\documentclass[useAMS,usenatbib]{mnras}

\usepackage{amsmath,amssymb,amsfonts}
\usepackage{bm}
\usepackage{graphicx}
\usepackage{color}
\usepackage[varg]{txfonts}
\usepackage{enumerate}
\usepackage{enumitem}
\usepackage{indentfirst}
\usepackage{float}

\begin{document}

\title[]{The effect of Large Magellanic Cloud on the satellite galaxy population in Milky Way analogous Galaxies}

\author[D. Zhang et al.] 
{Dali Zhang$^{1,2}$, Yu Luo$^{1}$\thanks{E-mail:luoyu@pmo.ac.cn}, Xi Kang$^{1,2}$\\
$^1$Purple Mountain Observatory, the Partner Group of MPI f\"ur Astronomie, 2 West Beijing Road, Nanjing 210008, China \\
$^2$School of Astronomy and Space Sciences, University of Science and Technology of China, Hefei 230026, China\\
}

\date{Manuscript Version: Jun 2019}

\pagerange{\pageref{2440}--\pageref{2448}} \pubyear{2019}

\maketitle

\label{firstpage}
 
\begin{abstract}
Observational work have shown that the two brightest satellite galaxies of the Milky Way (MW), the Large Magellanic
Cloud (LMC) and the Small Magellanic Cloud (SMC), are rare amongst MW analogues. 
It is then interesting to know whether the presence of massive satellite has any effect
on the whole satellite population in MW analogues. In this article, we investigate this
problem using a semi-analytical model combined with the Millennium-II Simulation. 
MW-analogous galaxies are defined  to have similar stellar mass or dark matter halo mass to the MW.
We find that, in the first case, the halo mass is larger and there are, on average, twice as many 
satellites in Milky Way analogs if there is a massive satellite galaxy in the system. This is mainly from the halo formation bias. The
difference is smaller if MW analogues are selected using halo mass. We also find that
the satellites distribution is slightly asymmetric, being more concentrated on the line connecting the
central galaxy and the massive satellite and that, on average, LMC have brought in
$14.7$ satellite galaxies with $M_{r}<0$ at its accretion, among which $4.5$ satellites are
still within a distance of $50$kpc from the LMC. Considering other satellites, we predict that there
are $7.8$ satellites with $50$kpc of the LMC. By comparing our model with the early data of
Satellites Around Galactic Analogs (SAGA), a survey to observe satellite galaxies around 100 Milky Way analogues, we find that SAGA has
more bright satellites and less faint satellites than our model predictions. A future comparison with the final SAGA data is needed.

\end{abstract}


%%%%%%%%%%%%%%%%%%%%%%%%%%%%%%%%%%%%%%%%%%%%%%%%%%%%%%%%%%%%%%%
\section{INTRODUCTION}
\label{sec:introduction}

The  cold  dark matter  (CDM)  model  makes  good predictions regarding structure formation on
large scales. However, on  small scales the  predictions are complicated, due  to the
complex  baryonic   processes.  Our  Milky   Way  (MW)  is   the  most
well-studied  galaxy in  the universe(e.g. \citet{Bla16}). 
In recent years, dozens of  new faint satellites of MW has been
observed  \citep[e.g.,][]{Kop15, Drilica15, Bec15},
therefore the  whole member of  MW satellites could be  predicted more
confidently  \citep[e.g.,][]{Trow13, New18}.  The
combination   of  satellite   spatial  distribution   and  kinematical
properties (e.g., \citet{McCon12}) makes  the MW to be  an excellent
local laboratory to constrain the  CDM model and baryonic physics. For
a  recent review  see  \citet{Bul17} and  \citet{Nav18}.

In  fact,  two well-known  problems  have  been found  when  comparing
observations with theory. One of them is that the satellites predicted in the
CDM  model are  much more numerous than  we have  observed (the  "missing
satellite"  problem, \citet{Kly99}; \citet{Moo99}). The
problem can  be solved by  suppressing the  star formation rate  in low-mass haloes \citep[e.g.,][]{Bul10, Mac10}. The other
problem  is that  the predicted  central mass  density is  higher than
observed  (the   "too-big-to-fail"  problem,  \citet{Boy12}. 
Several solutions  have been proposed, such as  the possibility that dark matter
is not cold \citep[e.g.,][]{Lov12, Vog13}, an
induced core arising from baryon physics  and star formation \citep[e.g.,][]{Bro14, Guo15, Wet16,
Bro17}, the MW  having  a  lower halo mass  \citep[e.g.,][]{Ver13, Die17} or the MW having a  particular
quirk of accreting its subhaloes (\citet{Kang16}).

To solve these problems, on the one hand major effort should be made to search for more faint MW satellites.
  On the other hand, it is  equally important to search
more  MW  analogues to identify  whether  the  MW is  an
outlier. A few studies have  shown that the MW  is not typical,
especially  in terms  of its distribution of bright satellites \citep[][]{Liu11, Guo11, Tol11, Str12}. 
Considering the brightness of Large and Small Magellanic Clouds
(LMC \&  SMC, with  $M_r=-18.6$ and  $M_r=-17.2$), studies  found that
11 percent of MW  analogues contain a satellite  like LMC  or SMC  and only
3.5 percent of  them contain equivalents to both bright satellites \citep{Liu11}. 
A similar  conclusion is also drawn using  simulation
data \citep{Boy10, Bus11, Rod13, Kang16}. 

Considering  the rarity  of LMC  like  satellite in  MW analogs,  one
important question is whether the presence of  a massive satellite,
such as the LMC, has any effect  on the number and spatial distribution of
the  full satellite  population.   In particular,  does  our MW  contain more or  fewer faint satellite galaxies than  its analogues?
If there are more satellites, how many of them are contributed by LMC?
Answering  these questions  will shed  light on  our search  for more  faint
satellites  in the  MW  and the comparison with  observations of MW
analogues. On the other hand, the  effect of a LMC-like  satellite on the
whole  satellite  population  may  depend  on how  we  define  the  MW-analogous galaxies. 
Usually, a MW-analogous galaxy is defined as a galaxy that has a dark
matter halo similar to that of the MW, around $10^{12}M_{\odot}$. It is well-known from the halo model (\citep{Zen05, Dea13, Mao15}) that the number of satellite galaxies, as well as their mass distribution, is related with halo assembly bias. \citet{Lu16} use an N-body simulation of MW-mass haloes with a fixed mass of $M_{vir}\sim 10^{12.1}M_\odot$ to investigate the realtion between LMC and other satellites. They found that the presence of the LMC is correlated with halo assembly history and a different assembly history will have an effect on the number of satellites. Although using dark matter halo to define a MW analogue can be easily applied in simulations, in an observational search of MW analogues it  is more straightforward to  use the stellar mass  or luminosity of the galaxy. The Satellites Aroud Galactic Analogs (SAGA) Survey is an ongoing  galaxy survey, the goal of which is to measure the distribution of satellite galaxies around 100 MW analogs  down to $M_r=-12.3$ \citep{Geh17}. In their  first phase of data, the survey has found eight MW analogues and made some comparisons between their samples and the Local Group.

In  this article,  we  investigate  the  effect of the brightest
satellite in the MW-like galaxies using a semi-analytical model of galaxy formation. Although halo assembly is the main driver for the variance of satellites population in the galaxy, some physical effects will produce additional scatter, such as cosmic reionization, galaxy mergers and tidal stripping. A semi-analytical model is able to include these physical process and our model (\citet{Luo16}) has acquired much better convergence for simulations with different resolutions, which is essential for modelling the faint satellites of the MW. The goal of this  work is to find  out in which definition  of MW-analogous galaxies the   presence  of  a   LMC-like  satellite  will   affect  the
distribution of the whole satellite population.  We introduce the data
and our definition of MW analogous galaxies in Section  2. In Section 3  we show the satellite numbers and spatial distribution.  In Section 4 we compare
the model predictions with the recent SAGA results on the distribution
of bright satellites to identify whether SAGA has missed some bright or
faint satellites.  We briefly discuss the results in Section 5.

%%%%%%%%%%%%%%%%%%%%%%%%%%%%%%%%%%%%%%%%%%%%%%%%%%%%%%%%%%%%%% 

\section{METHODS}

\begin{figure}
  \includegraphics[width=\linewidth]{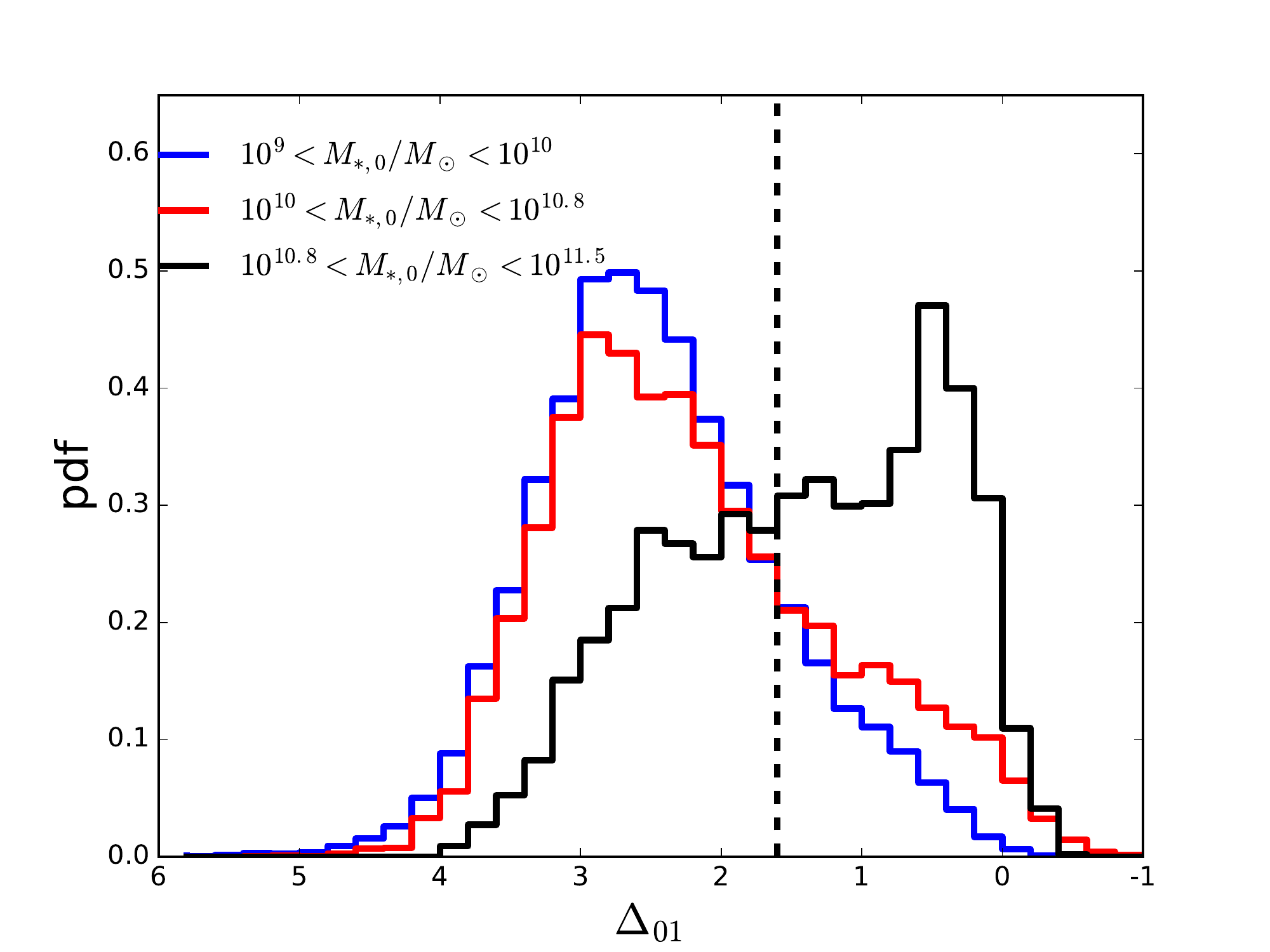}
  \caption{The distribution of $\Delta_{01}$, the gap between the stellar mass of central galaxy and the most massive satellite, for central galaxies in different mass bins. The black dash line is the value of $\Delta_{01}$ for the MW. }
  \label{fig:delta01}
\end{figure}

In  this  work,  we  use  the
semi-analytical galaxy formation model  of   \citet{Luo16}  implemented in the
Millennium-II simulation  (MS-II, \citet{Boy09}) to  produce  model
galaxies.  The original MS-II  simulation is  a dark  matter only  cosmological
simulation with $3072^{3}$  particles in a cube box with  each side 
$100Mpc/h$.  After rescaling the cosmological parameters from the WMAP1 
to the WMAP7 cosmology (a method developed by \citet{Ang10}), the box size of MS-II 
is rescaled from $100Mpc/h$ to $104.311Mpc/h$ and 
the mass of each particles are changed from $6.9\times10^{6}M_{\odot}/h$ to $8.5\times
10^{6}M_{\odot}/h$.  It is  shown  that such  a  simulation has  enough
resolution to  resolve faint satellite  down to  $M_{v} = -5$ \citep{Guo15}. 
The semi-analytical model of \citet{Luo16} is a resolution-independent model based on
the  Munich galaxy   formation  model: L-Galaxies \citep[e.g.][]{Kau93, Kau99, Spr01, Cro06, De07, Guo11, Guo13, Fu10, Fu13, Hen15}. 
For more detail  about the models, we refer the readers to the references above.

In  the   L-Galaxies  model,   galaxies  are  classified   into  three
types.  Type   0  galaxies  are   those  located  at  the   center  of
Friends-of-Friends (FoF)  halo groups and  we take these  galaxies as
centers (host galaxies)  of these groups. Therefore, there  is only one
type 0 galaxy in each FoF group and, in most cases, the type 0 galaxy
is the largest and the brightest member  of the group. Both type 1 and
Type 2  galaxies are regarded as  satellite galaxies in the  model. A
type 1  galaxy is  located at  the center  of a  subhalo, which  is an
overdensity within  the FoF halo \citep{Spr11}. The  halos /
subhaloes  contain  at  least  20  bound  particles  for  Millennium-II
\citep{Spr05, Boy09}. Type 2 represents
an  orphan galaxy  without a resolved  subhalo  and it  is usually  the
descendant of a Type 1 galaxy.

In the literature, the most common way to define MW-analogous galaxies is using
the dark matter halo mass. Great  efforts have been devoted to measure
the  dark matter  halo mass  of the  MW and  most results  lie between
$5\times  10^{11}M_{\odot}$ and  $3\times  10^{12}M_{\odot}$, with the majority 
concentrated  around  $10^{12}M_{\odot}$   (\citet{Li17};  see
\citet{Cal18} and \citet{Wang15} for a summary of different methods and results). 
In our work we use two selection criteria  to select MW-analogous galaxies.  The first one is  the virial
halo  mass and  we  select galaxies with  virial  mass between  $8\times
10^{11}M_{\odot}$ and $2\times 10^{12}M_{\odot}$  as MW analogues. The
second one is  to use the stellar  mass of the central  galaxy. In the
SAGA survey, \citet{Geh17} define MW-analogous galaxies using the K-band
luminosity of  the central  galaxy; they  obtained MW  analogues with
$-23  > M_{K}  > -24.6$  using the  abundance-match  method, which  is
slightly larger than  the reported K-band luminosity of the  MW in the
literature (e.g.,\citet{Kly02}). This luminosity  range 
corresponds roughly to stellar  mass  between  $2\times 10^{10}M_{\odot}$  and
$10^{11}M_{\odot}$.  In this  work, we  also use  this stellar  mass range to
select the MW-analogous galaxies, partly for comparison with the SAGA results.
For satellite galaxies, we use an $r$-band magnitude cut with $M_{r} < -6$ considering the resolution of the simulation. 
 
To  investigate the  effect of  a  massive satellite  galaxy on  the
distribution of  the whole galaxy  system, we  use the gap  in stellar
mass  between  the central  galaxy  and  the most  massive  satellite,
defined as $\Delta_{01} =  log_{10}(M_{*,0}/M_{*,1})$, to quantify the 
resemblance to the MW. Here $M_{*,0},M_{*,1}$ denote the stellar
mass of the central and the most massive satellite galaxy. 
The MW has
a  stellar mass  of  $\sim  6.5\times10^{10}M_\odot$  and  LMC  have  a
stellar mass  of $\sim  1.5\times 10^9M_\odot$,  thus we use $\Delta_{01} =1.6$ 
as a standard  to split our MW analogues into two classes,
with one class containing a large satellite while the other does not.

\begin{figure}
  \includegraphics[width=\linewidth]{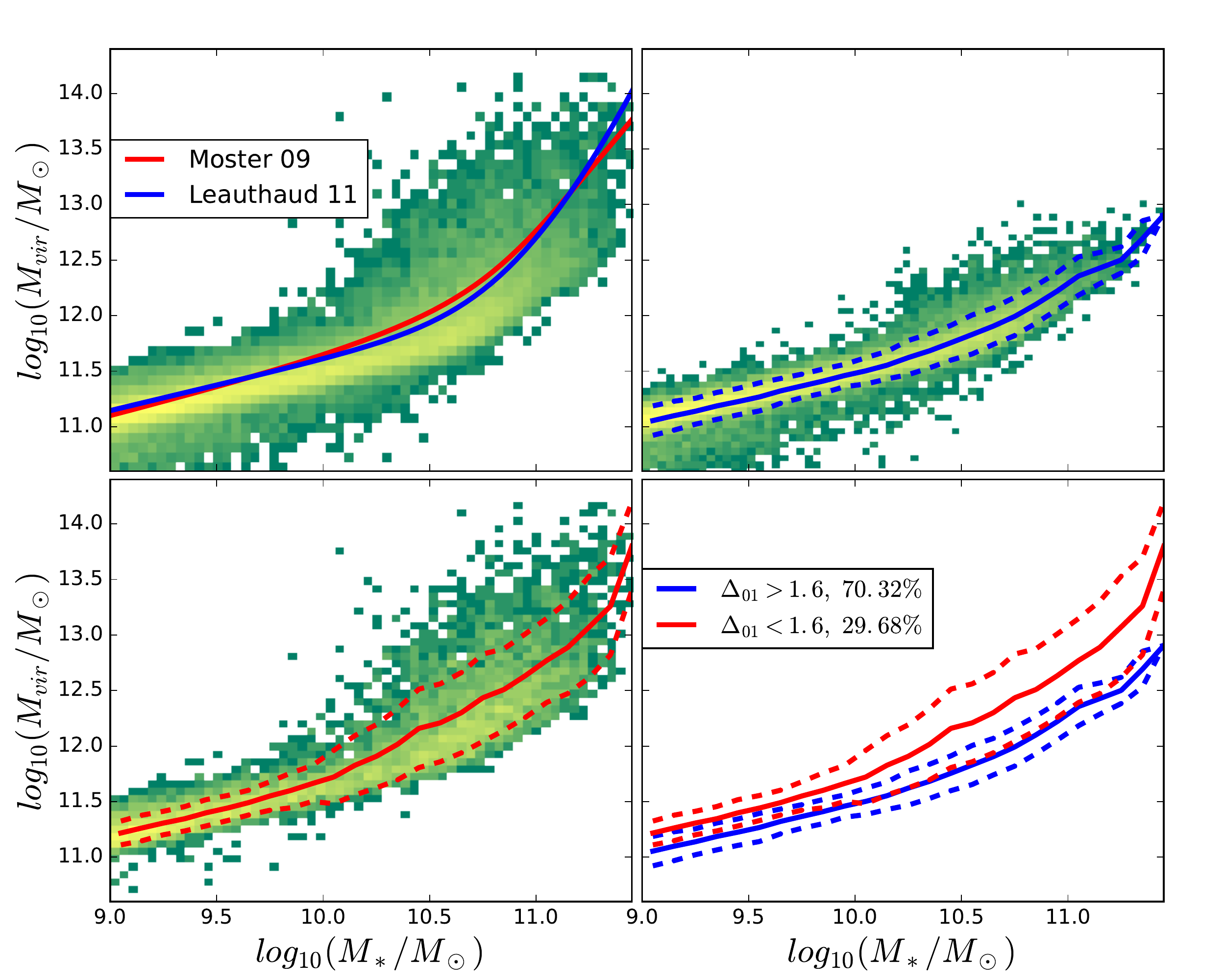}
  \caption{The stellar mass-halo mass distribution of center galaxies in our model. The upper left panel is the distribution of all galaxy samples in the model, along with the relations from the abundance matching by \citet{Lea11} and \citet{Mos10} The upper right panel is the distribution of galaxies with $\Delta_{01}>1.6$, and the blue solid and dashed lines repsent the median value and 1$\sigma$ deviation region. While the bottom left panel is the distribution of galaxies with $\Delta_{01}<1.6$. A comparsion of these two samples is shown in the bottom right panel.}
  \label{fig:sm-hm}
\end{figure}

%%%%%%%%%%%%%%%%%%%%%%%%%%%%%%%%%%%%%%%%%%%%%%%%%%%%%%%%%%%%%%%
\section{RESULTS}
\label{sec:result}

\subsection{how does the LMC affect the satellite mass distribution}

In this section, we investigate  the distribution of $\Delta_{01}$ and
its  impact   on  the  stellar  mass-halo   mass  relation,  satellite
number and spatial distribution.

We    first   show    the   distribution    of   $\Delta_{01}$    in
Fig.~\ref{fig:delta01}. We select central  galaxies in three mass bins
and  plot their  proportion distribution functions (PDFs)  of  $\Delta_{01}$. The  vertical  dashed line is the
$\Delta_{01}$ of the  MW. It is seen that for  small galaxies the gap
$\Delta_{01}$  is  larger,  while  for massive  galaxies  the  gap  is
smaller,  consistent  with  the results of  \citet{Kang16}. As  the
subhalo  mass function  (normalized by  the  host halo  mass) is  very
weakly  dependent on  the  host  halo mass  (e.g., \citet{Van06} 
), it is  interesting to ask why the gap  in stellar mass has a 
dependence on the central stellar mass  or the host halo mass. This is
mainly  because  the star  formation  in  low-mass galaxies  decreases
rapidly \citep{Guo10}, leading to a large gap  in stellar mass for a
given gap in  dark matter mass.  The plot shows  that for galaxies
with   mass  less   than the MW   ($10^{10.8}M_{\odot}$)  the   peak  of
$\Delta_{01}$ distribution is larger than  1.6, while for the galaxies
with stellar mass  larger than MW, the peak value  of $\Delta_{01}$ is
lower than  1.6.  It means that  whether a galaxy group contains  a LMC-like
satellite is correlated with the mass of the central galaxy.

In Fig.~\ref{fig:sm-hm}  we check the stellar  mass-halo mass relation
for central  galaxies in our model.   The upper left panel  is for all
central  galaxies  with  stellar mass  between  $10^{9}M_{\odot}$  and
$10^{11.5}M_{\odot}$ and the lines  show the relations from 
abundance  matching by \citet{Mos10} (2009, red line) and  
\citet{Lea11} (2011, blue line).  The upper right and lower left panels are 
for  central galaxies  with larger/smaller  gap. The  lower right  panel
compares the  median relations  for the two  samples. The  upper left
panel shows  that our semi-analtical model (SAM)  galaxies agree roughly with  the abundance-matching results,  
indicating that the model  parameters are correctly
tuned.  The  lower right  panel shows  that, at  given stellar  mass, a
galaxy with  lower $\Delta_{01}$  tends to have  a larger  virial halo
mass, which is easy to understand, since both $\Delta_{01}$ and halo mass are affected by halo assembly history. Such a trend is stronger in massive galaxies.

\begin{figure}
  \includegraphics[width=\linewidth]{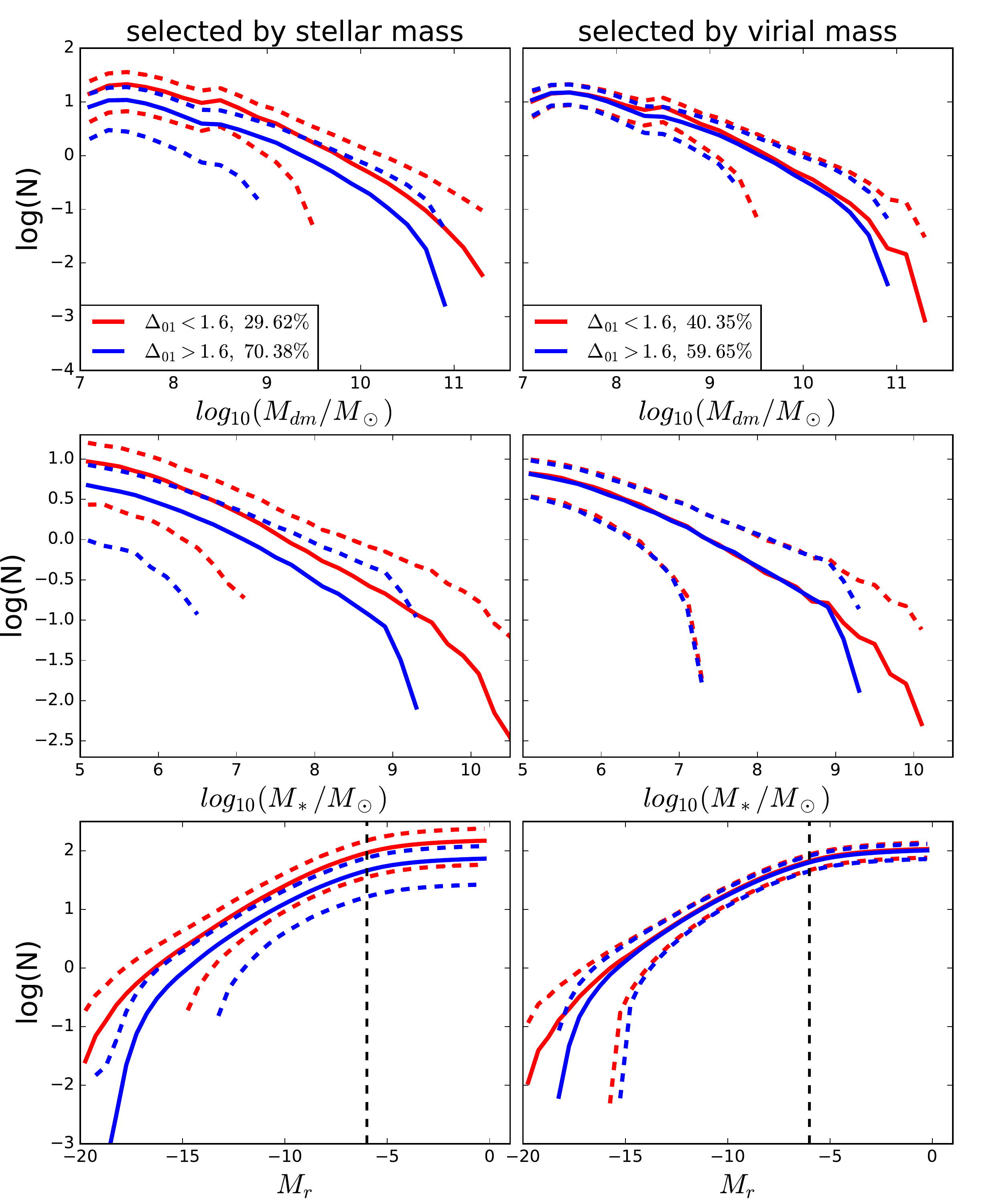}\\
  \caption{The satellite mass and luminosity distribution of MW analogues. Samples selected by stellar mass of central galaxy are shown in the left column, and  samples selected by halo virial mass are shown in the right column. The upper two rows are for satellite halo mass and stellar mass, the lowest one for satellite luminosity. Red lines represent samples with $\Delta_{01}<1.6$, blue lines samples with $\Delta_{01}>1.6$. Dashed lines represent 1$\sigma$ deviation. The vertical dashed lines ($M_{r} = -6$ ) in  the lower panel show the resloution limit of our model.}
  \label{fig:distribution}
\end{figure}

As the halo  virial mass of a  galaxy is a more  physical property, it
means that the presence of a massive satellite might have an effect on
the  whole  satellite population.  In  Fig.~\ref{fig:distribution}  we plot  the
dark matter mass  function, stellar  mass function  and luminosity
function of the satellites.  In the left column, galaxies are selected by stellar mass,
i.e, the  MW analogues are  selected with stellar  mass in the  range 
$2\times 10^{10}M_{\odot}$ - $10^{11}M_{\odot}$.  In the right column,
MW analogues  are selected using  halo virial  masses in the  range 
$8\times  10^{11}M_{\odot}$  -  $2\times 10^{12}M_{\odot}$.  In  each
panel, red lines are for galaxies with lower $\Delta_{01}$ and blue
ones for those  with larger $\Delta_{01}$. As this  selection is based
on a massive  satellite galaxy, which will then introduce  a bias at
the  massive  end of  each  panel,  we  do  not  count the most massive
satellite  galaxy in the sample with lower $\Delta_{01}$ and only  
show  the  mass  functions of  all  other satellite galaxies.

From Fig.~\ref{fig:distribution}, it is found  that, by selecting  MW
analogues using the halo virial mass,  satellite galaxy distributions
have no dependence  on $\Delta_{01}$ except at the very massive end: i.e., the presence  of a massive
satellite,  such as the LMC, has  no effect  on the  total and  luminosity
distribution of  other satellite galaxies.   However, if we  select MW
analogues using the stellar mass of  the central galaxy, the presence of
a massive  satellite will lead  to more satellite galaxies  around the
galaxy (left  column).  The left column  shows that the total  number of
satellite galaxies  will increase by a  factor of 2 compared with the cases
with  larger $\Delta_{01}$.  This is due mainly  to the  result that
galaxies with lower $\Delta_{01}$ have  larger halo mass. Such an effect
is significant and could be verified using a large survey  of MW analogues,
such as SAGA, with its full coverage of 100 galaxies. We will compare our model predictions briefly with the early results of SAGA in Sec.4.

\subsection{Satellite spatial distribution}

\begin{figure}
  \includegraphics[width=\linewidth]{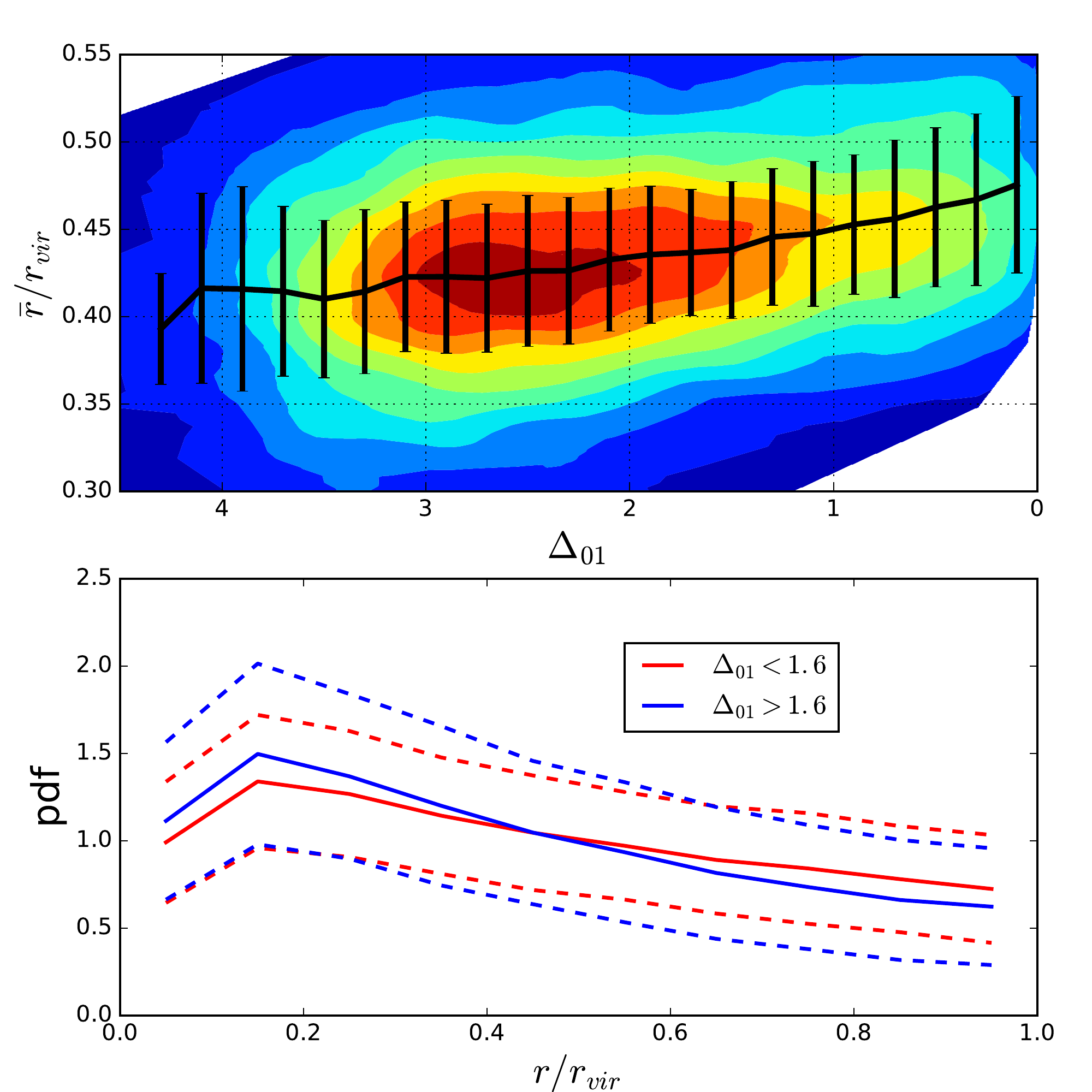}
  \caption{The central-satellites distance distribution with $M_{r,sate}<-6$. The upper panel is the distribution of average center-satellites distance with respect to $\Delta_{01}$, where the distance is normalized by the halo virial radius. The black line is the average value of the distribution, with error bars representing $ 1\sigma$ deviation. The bottom panel shows the PDF of the distance for two samples, where dashed lines represent the 1$\sigma$ deviation.}

  \label{fig:distance}
\end{figure}

In previous section, we find that the presence of a massive satellite will lead to more satellite galaxies around a galaxy selected by stellar mass. 
Here we investigate whether the radial and angular distribution of the satellites will be affected by the massive satellite.  We show the center-satellite distance distribution in Fig.~\ref{fig:distance}. In the upper panel, the contour shows the distribution, while the black line shows the average distance. In the bottom panel, we show the satellite distance distribution function in units of the virial radius. The dashed lines show the 1$\sigma$ variance. It is found that, when a galaxy system has a lower $\Delta_{01}$, more satellites are likely to stay slightly further from central galaxy and the average distance to the central galaxy is also larger. 

\begin{figure}
  \includegraphics[width=\linewidth]{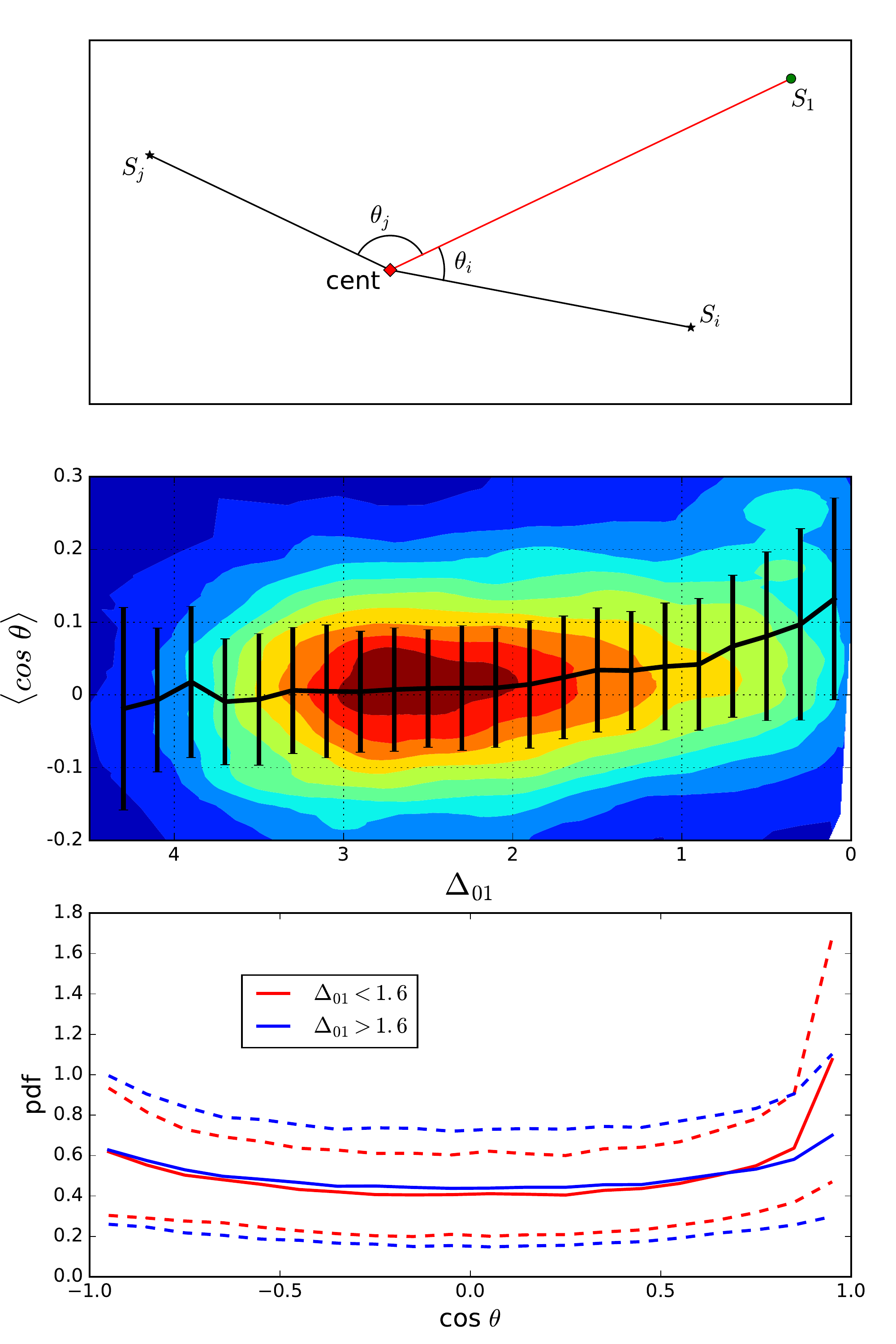}
  \caption{The angular distribution of satellites with $M_r<-6$ respect to the largest satellite galaxy. The upper panel illustrates the configuration of  angular position of satellites. Here, 'cent' denotes the center galaxy and $S_1$ represents the largest satellite, while $i,j\ge2$ means other satellites. In the middle panel, the contour plot represents the distribution of $\langle cos \theta\rangle$, with the average value of the contour and 1 $\sigma$ deviation shown by black line with error bars. The bottom panel shows the angular distribution of satellites in two samples.  }
  \label{fig:angular}
\end{figure}

We  also   show  the  angular   distribution  of  the   satellites  in
Fig.~\ref{fig:angular}.  We define  $\theta$ to  describe the  angular
position  of satellites with respect  to the  largest (ie. the most massive) 
satellite and  the
configuration of $\theta$ is shown  in the top panel. The distribution
of average value of cos$\theta$ is  shown in the middle panel.  As we
can  see, for  larger $\Delta_{01}$  value,  the average  value of  cos
$\theta$ is close  to $0$.  The average value of  cos$\theta$ tends to
be  larger  when  $\Delta_{01}$  decreases, which  means  the  angular
distribution of all  satellites  is affected by a large satellite,
with  a greater number  of  satellites located on the  same  side at  the
largest  satellite.  We  show  the angular  distribution of  two 
samples in  the bottom panel.  The angular  distribution of
satellites shows some kind of elliptical form, with more satellites staying at
the position with  cos$\theta \sim \pm 1$, especially for  galaxies with
lower $\Delta_{01}$.  For MW-like  samples, we limit $\Delta_{01}$
in  a narrow  range with  $1  < \Delta_{01}  <  2$, about  9 percent of  the
satellites stay  in a  narrow cone  that shares the  orientation of  the largest
satellite (cos $\theta>$ 0.9).

What  leads to  the slightly  larger  distance and  more narrow  angular
distribution  of  satellite  galaxies   if  there  is  a  massive
satellite,  such as  the LMC?  One possible  reason is  that the  most
massive  satellite is  recently accreted  and is  still orbiting in the
outer halo region  of the central galaxy. As a massive satellite galaxy
may  have  its own  satellit,  i.e.,  satellite of a satellite,  these
satellites are  still concentrated around the  massive satellite, 
leading to a  slight larger distance and  angular distribution towards
the  massive satellite.   Another possible  reason is  that there  is a  low
potential region between the center and  the largest satellites because of
their gravity, hence some of the satellites are trapped in the low potential
region.

\subsection{Satellites in neighborhood of LMC}

So far, more than 50 Milky Way satellites have been detected and 13 of them are locate within 50 kpc of the LMC \citep{Kop15, Bec15, Drilica15, Hom18, Kim15}. 
In this section, we will investigate the satellites in the neighborhood of LMC-like satellites and satellites of LMC
analogues before their accretion. In last sections, we selected the satellites in our model with a magnitude cut of $M_r<-6$, considering the resolution limit (see bottom panel in Fig.~\ref{fig:distribution}). However, in observations, 12 of 13 satellites in the neighborhood of LMC are faint with $-6<M_r<0$, except for the SMC ($M_r=-17.2$). In this section, we change our magnitude cut to $M_r<0$ to make a prediction of satellites near to the LMC. Here we note that our model predictions for very faint satellites may be affected by the simulation resolution.

In the CDM scenario, a satellite galaxy is the remnant of a merger event. For each halo in our model, we trace its massive satellite, labeled as $S_1$, back to the snapshot before it is accreted. In that snapshot, $S_1$ was a central galaxy in its own FoF group, with a number of satellites in that halo. After $S_1$ merged with a massive galaxy, its satellites become satellites of the new central galaxy and some of them have merged with the central, while others survived. 

\begin{figure}
  \includegraphics[width=\linewidth]{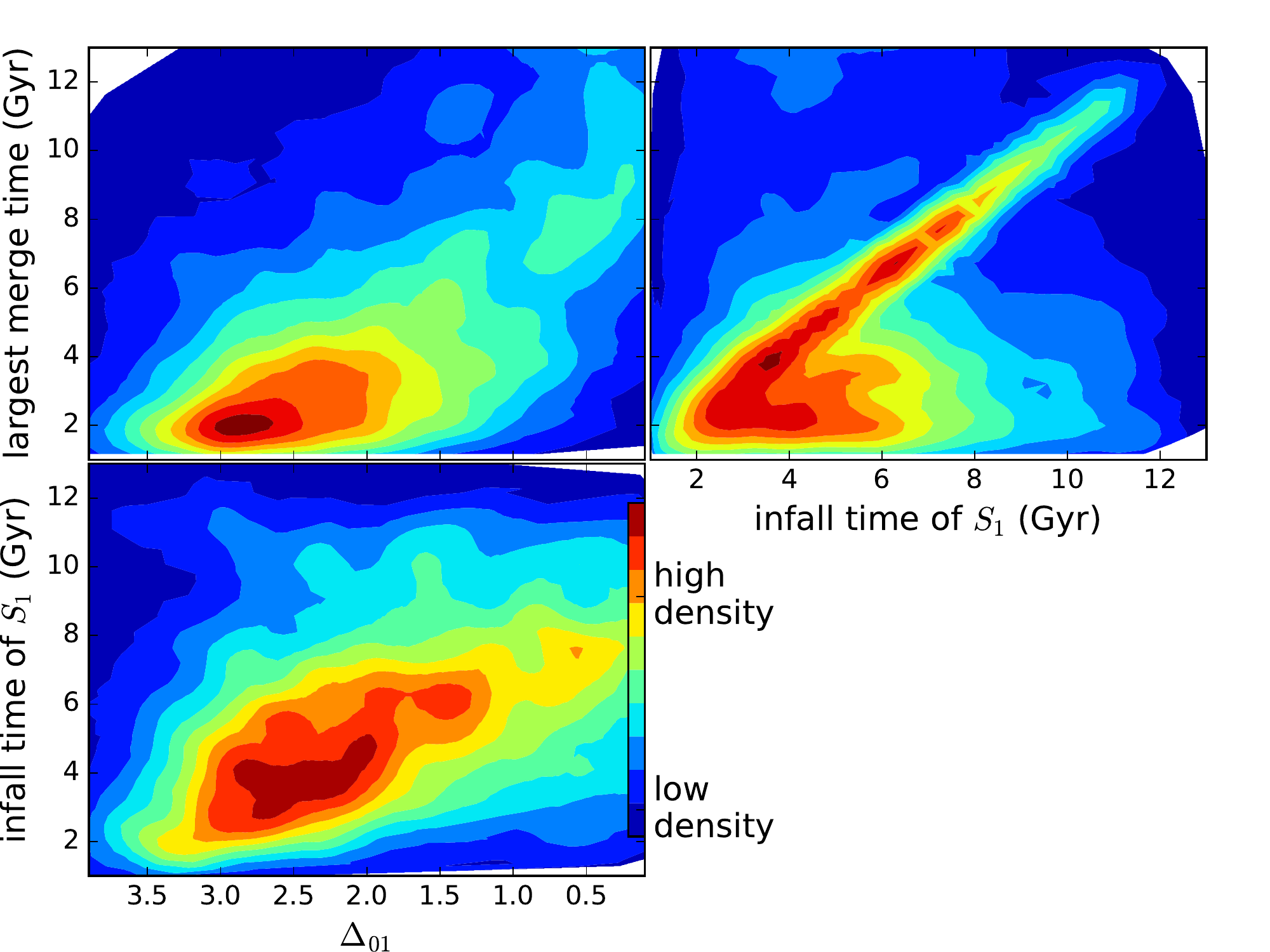}
  \caption{The relation between merger time and $\Delta_{01}$. The largest merger is defined as the time of the merger event that contributes the highest number of satellite galaxies. The upper left panel shows the distribution of the largest merger as a function of $\Delta_{01}$ and the bottom left panel shows the distribution of the infall time of $S_1$ as a function of $\Delta_{01}$. The merger times of both $S_1$ and largest merger present a correlation with $\Delta_{01}$. The right panel shows the relation between the time of largest merger and infall time of $S_1$, which indicates that in most cases the $S_1$ galaxy (most massive satellite) contributes the highest number of satellites during its accretion by the host galaxy.}
  \label{fig:merger}
\end{figure}

In this section, we investigate the infall time of $S_1$ and the fate of its satellite galaxies after $S_1$ is accreted. 
We also record the merger time, labelled as the largest merger time of the halo, in which the merger event contributed the greatst number of satellites. 
In Fig.~\ref{fig:merger}, we show the distributions of the infall time of $S_1$, the largest merger time and the gap $\Delta_{01}$. 
From the lower left panel, it is found that there is a correlation between infall time and $\Delta_{01}$, 
such that the satellite is more massive the later the merger event. 
This is reasonable, as the dynamical friction time is short for a massive satellite galaxy, so its presence indicates a very recent merger. 
The upper right panel shows that there is a very good correlation between the infall time of $S_1$ and the largest merger time, 
indicating that usually the most massive satellite contributes the highest number of satellite galaxies. 

The upper left panel is the relation between the largest merger time and $\Delta_{01}$. 
We use the largest merger time as a tracer to the formation time of the galaxy. 
When $\Delta_{01}$ is smaller, the largest merger time tends to be later, which means the $\Delta_{01}$ value 
of a galaxy is correlated with the history of halo accretion. 
Some investigates (e.g., \citet{Zen05}) have shown that the gap between the main halo and the largest halo is related to the halo assembly history, which is the main dirver for the gap in stellar mass between galaxies. However, there is some difference between the galaxy gap and the halo gap, since some other processes, such as cosmic reionization and galaxy mergers, might also make some contribution. We have shown in Fig~\ref{fig:sm-hm} that for the same stellar mass of the center galaxy, the halo mass tends to be larger when a galaxy has a smaller $\Delta_{01}$ value, suggesting that the deviation of halo mass is correlated with the halo assembly history. \citet{Lu16} have shown that that galaxies with a slower accretion ratio have a larger central galaxy, considering the relation between halo assembly history and $\Delta_{01}$.

Now we check the fate and origin of the satellites around $S_1$. We classify the satellite galaxies in each halo into three classes, $a,b,c$. Class $a$ represents all satellites which were satellites of $S_1$ before accretion and have survived until present, ie, they are still satellites in a galaxy group while not merged into other galaxies. Class $c$ are the neighbor satellite of $S_1$, where we define the neighbor as those within a distance of 50 kpc from $S_1$. Class $b$ are neighbors of $S_1$ and they were once satellites of $S_1$ before accretion.

\begin{figure}
  \includegraphics[width=\linewidth]{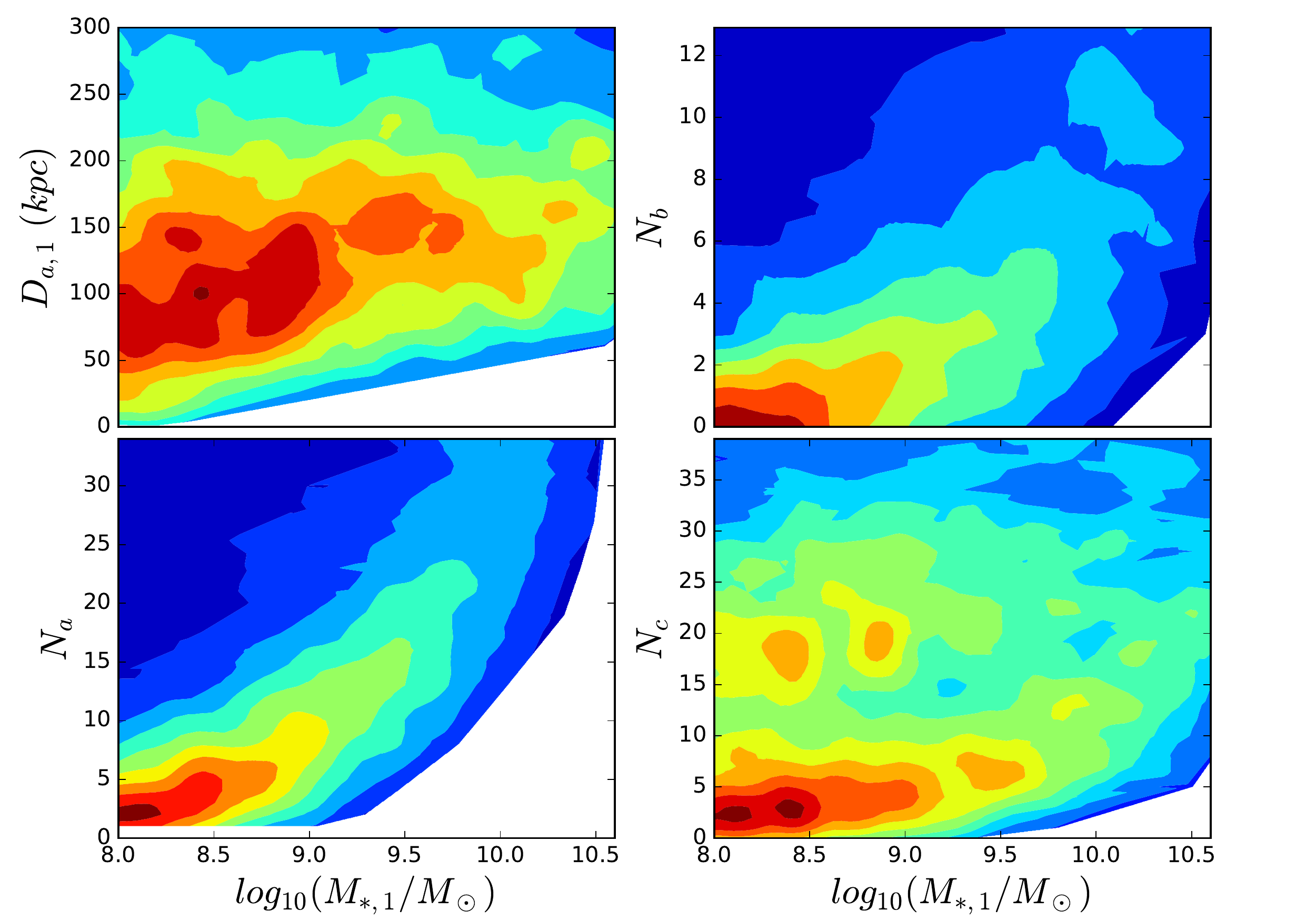}
  \caption{The distribution of satellites with $M_r<0$ affiliated with $S_1$ (the largest
satellite). Subscript \emph{a} represents a satellite-of-satellite, which means they were in a same FOF group as $S_1$ before $S_1$ was accreted by the host galaxy, subscript \emph{b} represents satellite-of-satellite in a neighbor (distanse $<$ 50 kpc) of $S_1$ and subscript \emph{c} represents neighbor of $S_1$. 
These satellites are represented by $S_a$ ($S_b$, $S_c$ respectively), while $N_a$ ($N_b$, $N_c$) represents the number of $S_a$ ($S_b$, $S_c$). Here $D_{a,1}$ means the average distance between $S_1$ and $S_a$. We find that $D_{a,1}$ tends to have a value $>$50 kpc, which means that most satellites of $S_1$ are escaped from its neighbor. Both $N_a$ and $N_b$ have good correlations with $M_{*,1}$, while $N_c$ shows no correlation with $M_{*,1}$.}
  \label{fig:classabc}
\end{figure}

In Fig.~\ref{fig:classabc} we show the properties of the satellites associated with $S_1$. Note that here the x-axis is the stellar mass of the massive satellite, labeled as $M_{*,1}$. The upper left panel shows the average distance between the largest satellite and class \emph{a} satellites. The average distance spread from dozens of kpc to hundreds of kpc and shows no significant correlation with the mass of the largest satellite. The average distances is of the order of the virial radius of the host halo, which means that, after infall, most of the satellites of $S_1$ escape from $S_1$ and become satellites of the center galaxy. The lower left and upper right panels show the number of class $a$ and class $b$ satellites, labeled as $N_a$ and $N_b$. They show that both $N_a$ and $N_b$ are well correlated with the mass $M_{*,1}$ of $S_1$, indicating that more massive $S_1$ satellites bring more satellites when they infall and may keep more of these satellites from their neighbors for a long time up to the present. The bottom right panel shows the distribution of the number of class $c$ satellites, $N_c$-$M_{*,1}$.  There is no strong correlation between $N_c$ and $M_{*,1}$, showing that the number of satellites in neighbors of $S_1$ cannot be predicted from the mass of $S_1$ only. 

Fig.~\ref{fig:classabc} is for all MW analogues with stellar mass between $2\times 10^{10}M_{\odot}$ and $10^{11}M_{\odot}$. If we  limit the LMC-analogues mass as $10^9<M_{*,1}/M_\odot<3\times10^9$, we find that there are on average $14.7^{+5.3}_{-5.7}$ satellites that were satellites of LMC before infall into the MW and $4.5^{+2.5}_{-2.5}$ satellites still stay in the neighborhood of the LMC, while others are stripped and distributed with a distance larger than $50$kpc to the LMC. If we use the satellite magnitude cut of $M_r=-6$ to select our samples, we find that there are on average $8.7^{+4.3}_{-3.7}$ satellites that were satellites of the LMC before their infall into the MW and $3.3^{+1.7}_{-2.3}$ satellites still stay in the neighborhood of LMC with $M_r<-6$. 
\citet{Doo17} and \citet{Shao18} also obtained similar conclusions with ours.

\begin{figure}
  \includegraphics[width=\linewidth]{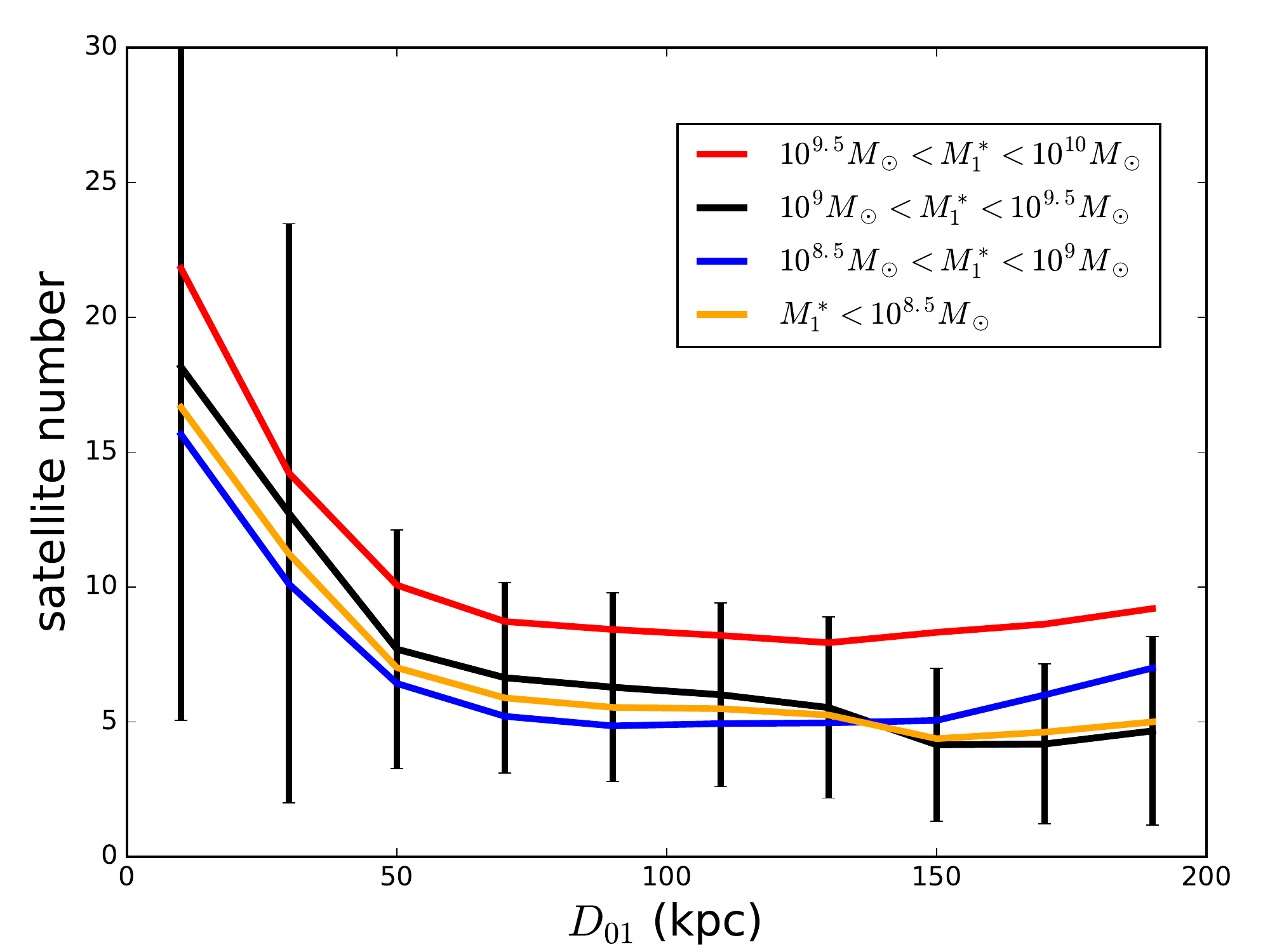}
  \caption{The relation between the number of neighbor ($N_c$) around $S_1$ and the distance
($D_{01}$) between central galaxy and the most massive satellite $S_1$. Here we show results for
$S_1$ in four mass bins. The black line with error bars shows the case for $S_1$ with mass similar to the LMC.}
  \label{fig:neighbor}
\end{figure}

To make a prediction about the neighbor satellites around the massive satellite $S_1$, in
Fig.~\ref{fig:neighbor} we show the relation between the number of satellites within $50$kpc of
$S_1$ and the distance between the massive satellite $S_1$ and the central galaxy, $D_{01}$.
Here we show the results of $M_{*,1}$ in different mass bins. It is shows that, for
$M_{*,1}$ larger than $10^{9.5}M_\odot$, the number of satellites is slightly larger than the
numbers in other mass bins, which means the mass of $S_1$ has some effect on the number of
neighbor satellites. For LMC like satellites (the black lines) or satellites smaller than LMC
(blue and orange lines) , the difference in the number of neighbor satellites is much smaller,
while the distance to the central galaxy has a dominant effect on the number of neighbor
satellites. We predict from this panel (black line) that $7.8^{+3.2}_{-3.8}$ satellites with $M_r<0$ should be observed within 50 kpc of the LMC. If we change the magnitude cut to $M_r<-6$, there are $5.4\pm3$ satellites. 

In reality, LMC has only one satellite with $M_r<-6$ (SMC, with $M_r=-17.2$) and 13 satellites with $M_r<0$ within $50$kpc. Compared with the model, there are more faint satellites and fewer bright satellites observed around the LMC.  One possible reason is that the resolution of the simulation we used, MS-II, is still too low and is not able to resolve and trace the very low mass haloes and subhaloes. In addition, in the model we do not consider the stripping of the satellite stellar mass, so the model may overestimate the number of bright satellites.

\section{COMPARISON WITH SAGA}
SAGA aims to measure the distribution of satellite galaxies up to the luminosity of Leo dwarf ($M_{r} < -12.3$) in around 100 MW analogs \citep{Geh17}. 
Early results have been published for eight galaxies.  
The SAGA team found that there is a large scatter between the eight galaxies and, in general, the observed satellite luminosity function has a flatter distribution compared with the prediction. Still, there is a missing satellite problem, with more faint satellites not being observed. In this section we compare the SAGA results with the predictions from our semi-analytical model. Note that here we select MW-analogous galaxies using the stellar mass of a central galaxy with the same coverage as SAGA, as introduced in Section.2.

\begin{figure}
  \includegraphics[width=\linewidth]{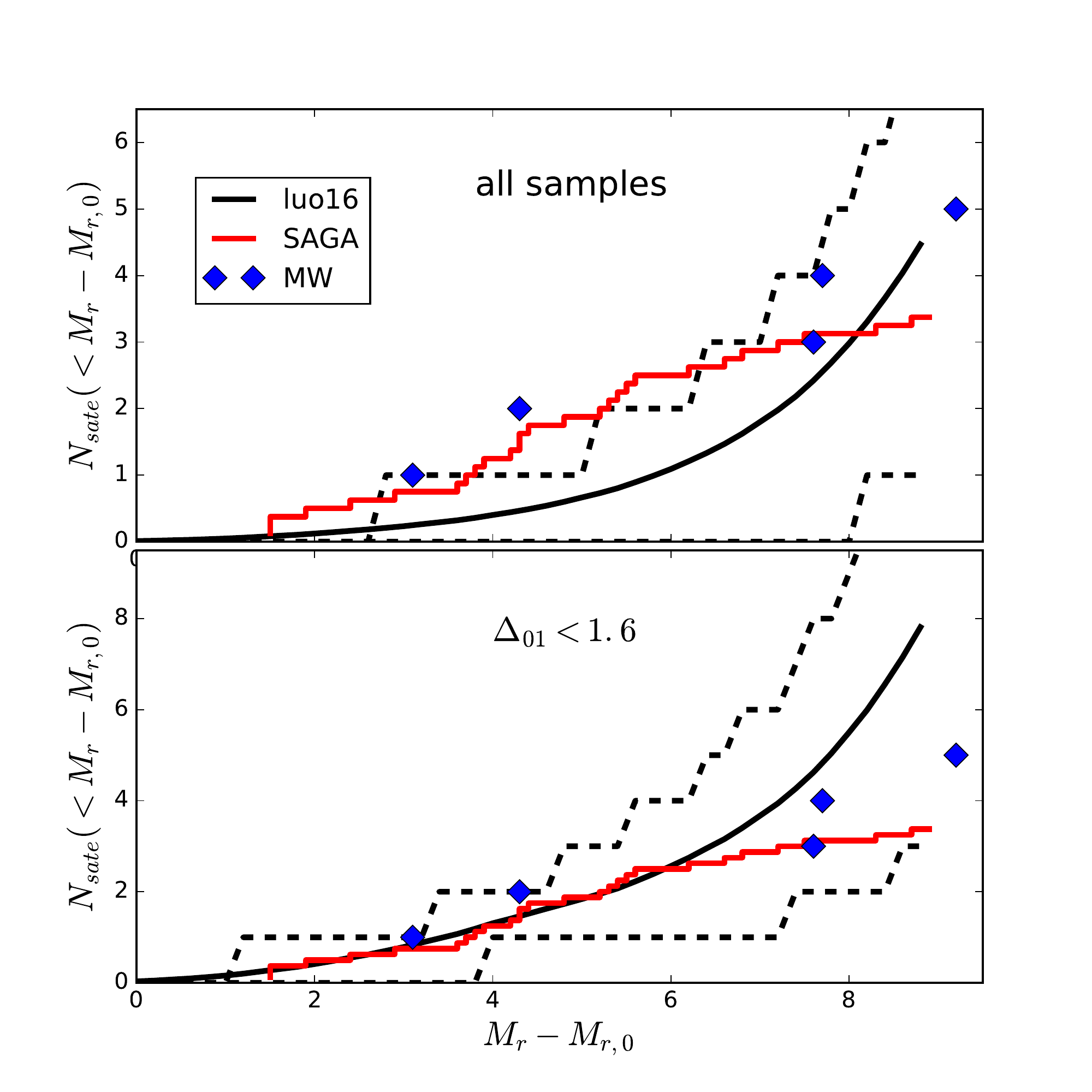}
  \caption{Satellite luminosity functions for observations and the model. The upper  panel is for
all central galaxies with stellar mass similar to the MW. The lower panel is for those central
galaxies containing a massive satellite. The black dashed lines show the $1 \sigma$ deviation of the
Luo16 model predictions. This shows that SAGA and the MW have a similar distribution of bright
satellites, but the MW has more faint satellites. The model contains more faint satellites than the
data, while the predicted number of bright satellites agrees with the data if the existence of a
massive satellite is considered (lower panel). }
  \label{fig:LFSAGA}
\end{figure}

In the first release data of SAGA, the eight galaxies have 27 satellites in total, among them, one galaxy has nine satellites, while four galaxies have only two satellites, thus the satellite luminosity function of each galaxy has large scatter and depends on the host luminosity. Here we combine the total satellite of the eight host galaxies and plot the luminosity function using the magnitude difference from the host. This will eliminate the dependence on the host luminosity. In Fig.~\ref{fig:LFSAGA} we show the satellite luminosity function as a function of $M_{r}-M_{r,0}$, where $M_r$, $M_{r,0}$ is the r-band luminosity of the satellite and the host galaxy. 

The upper panel of Fig.~\ref{fig:LFSAGA} shows that, in general, the satellite luminosity of the
MW is consistent with our model prediction, indicating that the SAM \citep{Luo16} has correctly
captured the physics of galaxy formation on small scales. This is not surprising, as the
model is based on the L-Galaxy model (e.g., \citet{Guo13}), which is tuned to fit the
galaxy stellar mass function up to the very faint end. The satellites distribution from SAGA is very
similar to that of the MW, except at the faint end, where SAGA has fewer faint satellites. Compared
with the SAM prediction, both SAGA and MW have more bright satellite galaxies, but SAGA has
fewer faint satellites than the predictions. One possible reason is that SAGA missed some faint satellites, which may be due to its observational strategy. On the other hand, our model might have something incorrect, which means some phenomenon in observations is 
not well reproduced, such as missing satellite problem. 

In the bottom panel of Fig.~\ref{fig:LFSAGA}, we plot the distribution of samples with
$\Delta_{01}<1.6$. We find that, once we compare SAGA and MW satellite distributions with 
samples that contain a large satellite, the model agrees well with observation data at the bright
end, while the model shows more satellites at the faint end. This indicates that the contribution of the most massive satellites should not be negelected when we define MW analogues using stellar
mass selection.

\begin{figure}
  \includegraphics[width=\linewidth]{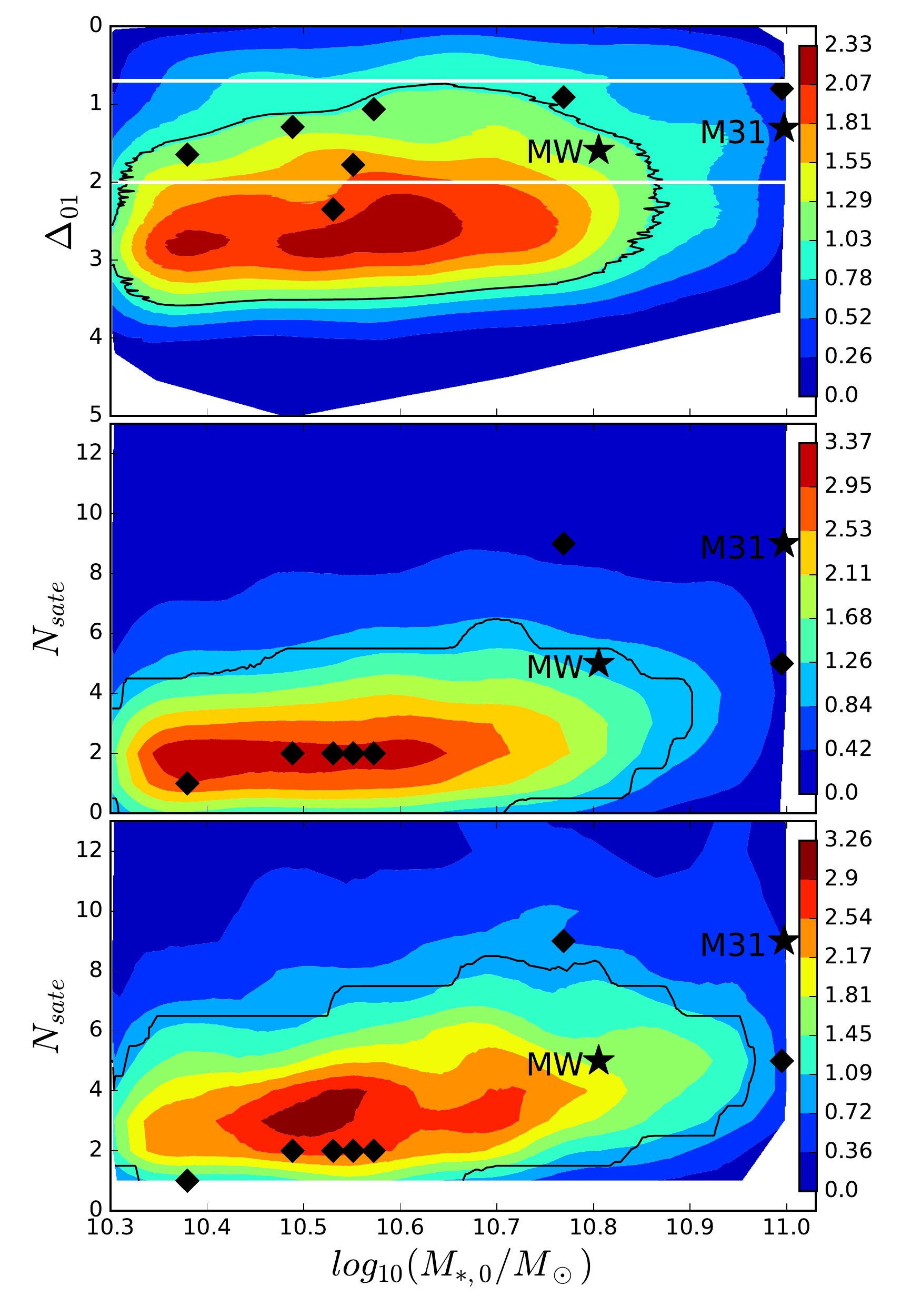}
  \caption{Upper panel: the relation between the mass gap ($\Delta_{01}$) and the stellar mass of
the central galaxy. Lower two  panels: the number of bright satellite galaxies (down to $M_{r} =
-12.3$). The color contours show the distribution of the model galaxy. Black diamonds
represent seven of the SAGA galaxies and black stars represent the MW and M31. The upper panel shows
that the SAGA galaxies and the MW have lower $\Delta_{01}$ compared with a model galaxy with similar
central stellar mass. However, considering the SAGA galaxies have lower $\Delta_{01}$, the
bottom panel shows the expected number of satellites from model galaxies selected with similar
stellar mass of the central galaxy and $\Delta_{01}$ (the region shown by the white lines in upper
panel). } 
  \label{fig:SAGAdelta01}
\end{figure}

Now we check further if the distribution of the brightest satellite in SAGA is consistent with the
model prediction. In upper panel of Fig.~\ref{fig:SAGAdelta01}, we show the relation between 
$\Delta_{01}$ and the stellar mass of the central galaxy. The black diamonds represent SAGA
galaxies, and black stars denote the MW and M31. It is seen that there is a weak trend that larger
central galaxies have smaller $\Delta_{01}$, consistent with the result in
Fig.~\ref{fig:delta01}. Compared with the model result, most SAGA have a smaller $\Delta_{01}$,
also consistent with that of MW and M31. From Section.3, we have found that the stellar
mass-halo mass relation is dependent on the stellar mass of the central galaxy and $\Delta_{01}$.

Then we estimate the deviation between SAGA and our simulation samples. 
Here we only count the number of satellites, $N_{r}$, with $M_{r}-M_{r,0} < 9$ according to the magnitude limit of SAGA. 
In the middle panel of Fig.~\ref{fig:SAGAdelta01}, the colored contour shows the distribution of
$N_{sate}$ for all model galaxies regardless of their $\Delta_{01}$. 
The SAGA galaxies seem to have similar satellite numbers to the model prediction, 
but we notice that all SAGA galaxies have smaller $\Delta_{01}$ than most of the model galaxies
(see the upper panel of Fig.~\ref{fig:SAGAdelta01}). 
Therefore, considering the effect of the most massive satellite, as discussed in Section 3, 
the SAGA galaxies may underestimate the number of satellites. 
We then select central galaxies covering the same stellar mass and $\Delta_{01}$ as the SAGA galaxies, 
corresponding roughly to the region covered by white lines in the upper panel of Fig.~\ref{fig:SAGAdelta01}. 
The expected number of satellites is shown in the lower panel of Fig.~\ref{fig:SAGAdelta01}. 
It is seen that both the satellite numbers of both the MW and M31 are consistent with our model, 
but the SAGA galaxy have a deficit of about two satellites on average per host galaxy. 
As previously shown, the satellites produced in our model that have not yet been found in observations should be faint, close to the detection limit of $M_{r} = -12.3$.

Finally, we note that SAGA currently only has satellite data for eight galaxies, as also mentioned by
their article \citep{Geh17}, i.e. the sample size is still too small. With more data available in
future, the satellite luminosity function can be measured more accurately and comparison with
the model will be more reliable.

%%%%%%%%%%%%%%%%%%%%%%%%%%%%%%%%%%%%%%%%%%%%%%%%%%%%%%%%%%%%%%%
\section{CONCLUSIONS AND DISCUSSION}
\label{sec:conclusion}

Both observational and theoretical studies have found that the Milky Way (MW) is atypical in that it
contains more bright satellites, such as the LMC and SMC, than the MW analogues. In this work, we use
model galaxies from a semi-analytical model combined with high-resolution N-body simulations to
study the effect of the largest satellite in a galaxy system on the whole satellite population.
In particular, we select the MW-analogous galaxies in our sample and investigate the
effect on the satellite number density and spatial distribution when there is an LMC-like
satellite. Our results can be summarized as follows:

\begin{itemize}
\item By selecting MW analogues using the stellar mass of the central galaxy, 
we find that galaxies with an LMC-like satellite have a larger dark matter halo mass and more satellite galaxies than those without an LMC-like satellite. 
The difference disappears when MW analogues are selected using the halo virial mass. 
 We also find that the gap between the center and the largest satellite is correlated with the largest merger time. 

\item The space distribution of satellites tends to be slightly away from the center galaxy and
to be asymmetric, with more satellites concentrated on the line between the central galaxy and the
largest satellite when there is a large satellite. The degree of anisotropy is correlated with
the mass of the largest satellite.

\item A large satellite is accreted into the MW more recently and brings more
satellites if its mass is larger. It is found that the LMC has brought $14.7^{+5.3}_{-5.7}$
satellites with $M_{r} < 0$ into the MW and about $4.5^{+2.5}_{-2.5}$
satellites still remain in the neighborhood with distance $<50$kpc from the LMC. Others have been
scattered into the MW halo. Considering the contribution of satellites not from the LMC,
we predict there are on average $7.8^{+3.2}_{-3.8}$ satellite galaxies within a distance of
$50$kpc from the LMC. This number is less than is observed around the LMC (13 satellites). We
note that the simulation we used does not have high enough resolution to resolve and trace very
low-mass haloes and subhaloes. Thus a further study using higher-resolution simulation is
called for.

\item When comparing the early data from SAGA with the Milky Way and the model, we find that SAGA has a 
similar distribution of bright satellites to the MW and our semi-analytical model and has fewer  
faint satellites than our model. On the one hand, we suggest that the deviation between SAGA and our model might be due to their observation 
techeniche, which means that SAGA is focusing on finding bright satellites. On the other hand, the deviation might be due to the inaccurate of our model. However, the conclusion should be taken with caution, 
as there are currently satellites data for eight galaxies from SAGA and a further comparison is needed
after it finishes the survey for about 100 MW analogues.

\end{itemize}

We select MW analogues containing an LMC-like satellite and predict the number of satellites
in the neighborhood of LMC-like satellites. In the true MW system, LMC and SMC stay very
close to each other and each of them has their own small-scale satellite system. LMC and SMC
might affect each other and the satellites around them may be influenced by these two big
members, which might make the satellite distribution deviate from our prediction. Besides this,
such as the shape and colour of the central galaxy and the large-scale
environment, might also influence the satellite distribution; these are not included in this
work. We believe that further high-resolution hydrodynamical simulation of MW analogues would be ideal
to study the satellites around the LMC and SMC.

\section*{Acknowledgments}
The Millennium-II Simulation data bases used in this article and the web application providing online access to them were constructed as part of the activities of the German Astrophysical Virtual Observatory (GAVO). We thank the anonymous referee for valuable suggestions on the article.
This work is supported by the National Key Basic Research Program of China (2015CB857003), the NSFC (No. 11333008, 11825303, 11861131006, 11703091).
We thank Qi Guo and the members of our labotory for useful discussion.

\end{document}